# In search of invariants for viscous liquids in the density scaling regime: Investigations of dynamic and thermodynamic moduli

Agnieszka Jedrzejowska,[1,2] Andrzej Grzybowski,[1,2,*] and Marian Paluch[1,2]

[1]*Institute of Physics, University of Silesia in Katowice, Uniwersytecka 4, 40-007 Katowice, Poland*

[2]*Silesian Center for Education and Interdisciplinary Research, 75 Pułku Piechoty 1, 41-500 Chorzów, Poland*

** Corresponding authors' email: andrzej.grzybowski@us.edu.pl*

## Abstract

In this paper, we report on nontrivial results of our investigations of dynamic and thermodynamic moduli in search of invariants for viscous liquids in the density scaling regime by using selected supercooled van der Waals liquids as representative materials. Previously, the dynamic modulus $M_{p\text{-}T}$ (defined in the pressure-temperature representation by the ratio of isobaric activation energy and activation volume) as well as the ratio $B_T/M_{p\text{-}T}$ (where $B_T$ is the thermodynamic modulus defined as the inverse isothermal compressibility) have been suggested as some kind of material constants. We have established that they are not valid in the explored wide range of temperatures $T$ over dozen decades of structural relaxation times $\tau$. The temperature dependences of $M_{p\text{-}T}$ and $B_T/M_{p\text{-}T}$ have been elucidated by comparison with the well-known measure of the relative contribution of temperature and density fluctuations to molecular dynamics near the glass transition, i.e., the ratio of the isochoric and isobaric activation energies, $E_V^{act}/E_p^{act}$. Then, we have implemented an idea to transform the definition of the dynamic modulus $M_{p\text{-}T}$ from the p-T representation to the V-T one. This idea relied on the disentanglement of combined temperature and density fluctuations involved in isobaric parameters has resulted in finding an invariant for viscous liquids in the density scaling regime, which is the ratio of the thermodynamic and dynamic moduli, $B_T/M_{V\text{-}T}$. In this way, we have constituted a characteristic of thermodynamics and molecular dynamics, which remains unchanged in the supercooled liquid state for a given material, the molecular dynamics of which obeys the power density scaling law.

## Introduction

In the condensed matter physics, a lot of effort has been put into searching material constants that characterize both thermodynamic and dynamic properties of various materials. Especially in case of the thermodynamically metastable supercooled liquid state, the nature of which is still shrouded in mystery, there is an urgent need for determining physically well-defined invariants, which would let us gain a better insight into the relevant factors that affect molecular dynamics near the liquid-glass transition.

In the last decade, a crucial role of the interplay between temperature and density fluctuations has been recognized in molecular dynamics of glass-forming liquids [1,2]. A few measures have been suggested to quantify the relative influence of the fluctuations of temperature ($T$) and density ($\rho$ equivalent to the inverse specific volume $V^{-1}$) on molecular dynamics in the vicinity of the glass transition. Among them the most popular one is probably the ratio of the isochoric and isobaric activation energies [3], $E_V^{act}/E_p^{act}$ , where the activation quantities are defined here by using a time scale $\tau$ of molecular dynamics as follows [4]

$$E_V^{act} = R\left(\partial \ln \tau / \partial (1/T)\right)_V \tag{1}$$

$$E_p^{act} = R\left(\partial \ln \tau / \partial (1/T)\right)_p \tag{2}$$

with the gas constant $R$. In case of many nonionic and nonpolymeric glass formers, the dynamic time scale $\tau$ typically means the structural relaxation time. For polymers, it is usually the segmental relaxation time, and for ionic liquids, it can be the ionic conductivity relaxation time. Independently of material class, the activation energies $E_p^{act}$ and $E_V^{act}$ can be also defined by using other dynamic quantities such as viscosity $\eta$ and ionic dc-conductivity $\sigma_{dc}$ instead of the dynamic time scale $\tau$. An undoubted advantage of the measure $E_V^{act}/E_p^{act}$ is its limited value ranging between two extreme ideal dynamic cases, i.e., *0* for molecular dynamics governed only by so-called *free volume* in a system and *1* for purely thermally activated processes. However, it has turned out that the ratio $E_V^{act}/E_p^{act}$ is not a material constant independent of thermodynamic conditions. For instance, the value of $E_V^{act}/E_p^{act}$ increases with increasing pressure in isochronal conditions determined by a constant structural relaxation time ($\tau$=*const*) [5]. This finding has been especially discussed by using an equivalent representation of the measure $E_V^{act}/E_p^{act}$ via the ratio of the isochoric and isobaric fragility parameters [6,7,8,9,10,11,12,13,14], $m_V = \left(\partial \log_{10}\tau / \partial(T_g/T)\right)_{V,T=T_g}$ and $m_p = \left(\partial \log_{10}\tau / \partial(T_g/T)\right)_{p,T=T_g}$, which yields $m_V/m_p = E_V^{act}/E_p^{act}$ at the glass transition temperature $T_g$, but it can be also considered at any constant relaxation time $\tau$ by replacing $T_g$ in the definitions $m_V$ and $m_p$ with $T_\tau$, i.e., $T_\tau = T$ at which a system is characterized by the relaxation time $\tau$. The isobaric fragility parameter has been widely propagated by Angell [15,16] as a relevant characteristic of glass-forming liquids. The other definitions of the fragility parameter in the isochoric and isothermal conditions have become very useful to analyze the experimental relaxation data measured e.g. by means of the broadband dielectric spectroscopy (BDS) not only at ambient pressure but also in high pressure conditions. The BDS measurements typically carried out in isobaric or/and isothermal conditions were initially described in the temperature-pressure (*T*-*p*) domain. In the last decade, the power law density scaling idea (according to which e.g. the structural relaxation times $\tau$ experimentally determined along different isobars and isotherms can be plotted onto one master curve as a function $\tau = f(TV^\gamma)$, where the scaling exponent $\gamma$ is a material constant independent of thermodynamic conditions) has given evidences that different analyses of such high pressure experimental data can provide more advantages by transforming them into the temperature-volume (*T*-*V*) domain [1,2]. For instance, in case of various materials that belong to different material groups (including supercooled van der Waals and ionic liquids as well as polymer melts), which obey the power density scaling law, some rules have been formulated for the fragility parameters defined in different thermodynamic conditions (i.e., *p=const*, *V=const*, and *T=const*) [17]. Consequently, the pressure behavior of the ratio, $m_V / m_p = E_V^{act} / E_p^{act}$, has been confirmed in the density scaling regime as increasing with increasing pressure (at $T=T_g$ or $\tau$=*const*), because $m_V = const$ and $m_p$ decreases with increasing pressure in this case.

Recently, an alternative approach has been suggested by Ingram et al. [18,19,20] who have considered the ratio of the isobaric activation energy $E_p^{act}$ and the activation volume $V_T^{act}$ defined as follows [4],

$$V_T^{act} = RT\left(\partial \ln \tau / \partial p\right)_T , \tag{3}$$

as a promising quantity that can be used to classify various ionic systems. The authors showed that values of the so-called dynamic modulus,

$$M_{p-T} = E_p^{act} / V_T^{act} \tag{4}$$

considerably differ between polymeric electrolytes (1.1GPa ≤ $M$ ≤ 2.4GPa) and molten salts ($M$ > 5GPa) [18,19]. Moreover, the values of $M_{p-T}$ seemed to be independent of different dynamic processes exploited to evaluate $E_p^{act}$ and $V_T^{act}$ for a given material. For instance, the authors found no significant differences between the values of the dynamic modulus received from both the analyses of the ionic conductivity and structural relaxation times of a molten salt silver iodomolybdate [20]. In addition, based on the analysis of experimental data in narrow temperature ranges, they suggested that $M_{p-T}$ can be determined for a given material from a linear correlation between $E_p^{act}$ and $V_T^{act}$ as its constant slope coefficient to a good approximation. Therefore, they postulated that $M_{p-T}$ can be used to characterize dynamic properties of various materials. What is more, taking into account the dynamic modulus as some kind of 'material constant', Ingram et al. suggested [18] a straightforward relation between $M_{p\text{-}T}$ and a parameter characterizing thermodynamic properties of materials, i.e., the isothermal bulk modulus $B_T$ defined as the inverse isothermal compressibility $\kappa_T^{-1}$,

$$B_T = -(\partial \ln V / \partial p)_T^{-1} \tag{5}$$

However, the proportionality between the dynamic and thermodynamic moduli, $M_{p\text{-}T} \sim B_T$, potentially providing a tempting linkage between molecular dynamics and thermodynamics of glass formers, were argued only phenomenologically by showing the same order of magnitudes of the moduli $M_{p\text{-}T}$ and $B_T$ in tested ionic systems by the authors [18-20].

In our previous work [21], using mainly dc-conductivity data collected for aprotic ionic liquid [$C_8$MIM][$NTf_2$] under isothermal and isobaric conditions, we have tested the hypotheses proposed by Ingram et al. Our findings have turned out to be model dependent. In those previous investigations of the dynamic modulus $M_{p\text{-}T}$ and its relation to the isothermal bulk modulus $B_T$, we have employed both the T-p and T-V versions [22,23,24,25] of the Avramov entropic model applied to describe the thermodynamic evolution of $\sigma_{dc}$. While we have obtained ambiguous results based on the T-p version of the Avramov model, its T-V version has brought a more reliable and interesting outcome. We have established a variability of the values of $M_{p\text{-}T}$ with temperature changes, which mimics the temperature dependence of $B_T$ determined from pVT data, but only to some extent, because the moduli $M_{p\text{-}T}$ and $B_T$ satisfy a more complex relationship than that suggested by Ingram et al.

In this paper, we focus on supercooled van der Waals liquids to study the dynamic modulus $M_{p\text{-}T}$ and its relation to the thermodynamic modulus $B_T$ in extremely broad ranges of dielectric structural relaxation times (10-12 decades of $\tau$) determined in wide temperature ranges. In this way, the hypotheses suggested by Ingram et al. are to be thoroughly verified in case of the prototypical glass-forming materials, the molecular dynamics of which is controlled by the structural relaxation process and obeys the power density scaling law. Nevertheless, our main goal is to answer the intriguing question *whether or not there is a relevant parameter characterizing both the molecular dynamics and thermodynamics of supercooled liquids, which is material-dependent, but invariant for a given material in the wide dynamic and thermodynamic range*.

**Results and discussion**

In order to verify the thermodynamic evolution of the dynamic modulus $M_{p\text{-}T}$ defined by Eq. (4), and its relation to the thermodynamic modulus $B_T$, we have selected three well-known van der Waals liquids, such as 1,1'-bis(p-methoxyphenyl)cyclohexane (BMPC), phenylphtalein-dimethylether (PDE), and propylene carbonate (PC), which were measured by means of the BDS technique in their supercooled state at ambient pressure in the range of dozen decades of their structural relaxation times $\tau$ and also at high pressure in some narrower ranges of $\tau$. These dielectric data earlier reported in literature densely cover broad temperature ranges especially at ambient pressure, i.e., 242K – 320K for BMPC [26,27] 295K – 415K for PDE [28,29], and 159K – 371K for PC [30]. However, there are not such rich data sets collected in high pressure conditions in which the dielectric measurements are more complex and have more limitations than those at ambient pressure. Therefore, to thoroughly examine the values of the dynamic modulus $M_{p\text{-}T}$ defined by Eq. (4), which involves the isobaric activation energies calculated from Eq. (2) along an isobar of pressure $p$ at given temperatures $T_i$ and the corresponding activation volumes evaluated from Eq. (3) at pressure $p$ along the isotherms of the temperatures $T_i$ , we have applied the criterion for the power law density scaling [31],

$$T_\tau V_\tau^\gamma = const \text{ at } \tau = const, \tag{6}$$

to generate high pressure isotherms of structural relaxation times $\tau$ at each temperature at which $\tau$ were determined from the BDS measurements at ambient pressure. Then, the values of the dynamic modulus $M_{p\text{-}T}$ could be numerically calculated by us according to Eq. (4) at each temperature at which $\tau$ were determined from the BDS measurements at ambient pressure. It is worth noting that the density scaling criterion is usually also valid along the $T_g$-line and at any $\tau = const$ to a good approximation for real materials independently of applied units of $\tau$ (i.e., not only reduced units of the isomorph theory [32,33] but also simply seconds) [34,35,36]. It should be emphasized that the suggested procedure is not only very useful, but also allows us to avoid using any other models to approximate the temperature-pressure dependences $\tau(T, p)$ established from the BDS measurements. Thus, this procedure offers some kind of numerical experiment, the results of which enable to supplement the measurement data if the density scaling criterion given by Eq. (6) is satisfied to a good approximation at each $\tau$ in the analysis range. It is reasonable to employ Eq. (6) in case of all the tested supercooled liquids, because each of them very well obeys the power density scaling law with the scaling exponent $\gamma$ being a material constant independent of thermodynamic conditions as shown in literature with $\gamma$=8.0 for BMPC [37], $\gamma$=4.42 for PDE [37], and $\gamma$=3.7 for PC [30]. To determine the volume $V$ as a function of $T$ and $p$ in Eq. (6), we have also employed $pVT$ measurement data and their parametrizations in terms of the equations of state (EOSs), which were used to find the values of the scaling exponent $\gamma$ for these materials. Thus, we have exploited the Tait EOS [38] parameterized for PC [30] and another EOS recently derived [39,40,41] for supercooled liquids and parametrized for PDE [39] and BMPC [42].

The aforementioned method based on the density scaling criterion given by Eq. (6) has enabled us to generate sufficiently dense data sets for the isothermal dependences $\tau(p)$ at all temperatures along the experimentally determined isobaric dependence $\tau(T)$ at ambient pressure $p_0$ to find reliable values of the activation quantities $E_{p_0}^{act}(T)$ and $V_T^{act}(p_0)$ by numerical differentiations according to the definitions given by Eq. (2) and Eq. (3). Then, the values of the dynamic modulus $M$ have been accurately evaluated from Eq. (4) at ambient pressure in the wide temperature ranges within which structural relaxation dynamics of the investigated supercooled van der Waals liquids significantly changes, i.e., the structural relaxation times $\tau$ increase by about 11 decades for all examined glass formers, while $T$ decreases by 78K for BMPC, 120K for PDE, and 212K for PC. As can be seen in Fig. 1, the evaluated values of the dynamic modulus $M_{p\text{-}T}$ are not constant for a given material at ambient pressure, but increase in a continuous manner with decreasing temperature. For BMPC and PDE, the changes in the values of the dynamic modulus are moderate in the considered temperature ranges, i..e., they are at the level of 26% and 29%, respectively. However, the values of $M_{p\text{-}T}$ change even by 114% in case of PC, the dielectric measurements of which have been performed in a considerably wider temperature range at ambient pressure than those carried out on BMPC and PDE.

In the next step of our analysis of the representatives of supercooled van der Waals liquids, we have calculated values of the isothermal bulk modulus $B_T$ from its definition expressed by Eq. (5) by using the same EOSs and their parametrizations of the dependences $V(T,p)$ as those applied to the density scaling criterion given by Eq. (6). As can be seen from the lower insets in the panels in Fig. 1, the coefficient values of the proportionality suggested by Ingram et al., $M_{p\text{-}T} \sim B_T$, are not constant for a given material, because the values of the ratio of the thermodynamic and dynamic moduli, $B_T/M_{p\text{-}T}$, increase with decreasing temperature by about 15% for BMPC, 32% for PDE, and 65% for PC.

Thus, neither $M_{p\text{-}T}$ nor $B_T/M_{p\text{-}T}$ are any invariant parameter characterizing the molecular dynamics or/and the linkage between molecular dynamics and thermodynamics of the selected van der Waals liquids tested in the supercooled state in the wide temperature range at ambient pressure. To better understand the temperature effect on the moduli, we have compared them with the well-known measure of the relative contribution of temperature and density fluctuations to molecular dynamics of supercooled liquids, which is the ratio $E_V^{act}/E_p^{act}$. It is worth noting that the procedure based on Eq. (6), which has been used to generate a series of isotherms to calculate the activation volume, can be also applied to generate an analogous series of isochores at volumes $V_0(T)=V(T,p_0)$ found from the appropriate EOS at temperatures $T$ at which the structural relaxation times were determined from the BDS measurements at ambient pressure $p_0$. Then, numerically differentiating in terms of the definition given by Eq. (1), we have been able to accurately evaluate values of the isochoric activation energy, $E_{V_0(T)}^{act}(T)$, which have been divided by the corresponding values of the isobaric activation energy, $E_{p_0}^{act}(T)$, to establish reliable values of the ratio $E_V^{act}/E_p^{act}$ along the isobar $\tau(T)$ at ambient pressure. From the inspection of Fig. 1, one can note that both $M_{p\text{-}T}$ and $B_T/M_{p\text{-}T}$ behave qualitatively in the same way as the ratio $E_V^{act}/E_p^{act}$ with changing temperature, i.e., all the quantities increase with decreasing $T$. This finding is worthy of a detailed further discussion.

Comparing the ratio $E_V^{act}/E_p^{act}$ and the dynamic modulus $M_{p-T}=E_p^{act}/V_T^{act}$, at first glance, one can state that the former is defined by using some *isolines* in the *V-p* representation, i.e., an isochore and an isobar at their intersection point, while the latter definition involves some isolines in the *p-T* representation, i.e., an isobar and an isotherm at their intersection point. Taking into account the definitions of the activation quantities, given by Eqs. (1) – (3), one can draw a few important conclusions about the physical meaning of $E_V^{act}$, $E_p^{act}$, and $V_T^{act}$. That is one can see that $E_V^{act}$ and $V_T^{act}$ are the parameters, which *well* *distinguish* the effect of changing temperature fluctuations on molecular dynamics at a fixed level of density fluctuations along an isochore (at *V*=*const*) and the effect of changing density fluctuations on molecular dynamics at a fixed level of temperature fluctuations along an isotherm (at *T*=*const*), respectively. However, $E_p^{act}$ involves a combined effect of changing temperature and density fluctuations along an isobar (at *p*=*const*). Therefore, the ratio $E_V^{act}/E_p^{act}$ quantifies the relative contribution of temperature fluctuations to the overall fluctuations of *T* and $\rho$. On the other hand, the ratio $V_T^{act}/E_p^{act}$ is a supplementary measure that characterizes the relative contribution of density fluctuations to the overall fluctuations of *T* and $\rho$. Thus, the increasing role of the temperature fluctuations in molecular dynamics with decreasing *T* along the atmospheric isobar, which is suggested by the established increasing values $E_V^{act}/E_p^{act}$ on cooling at ambient pressure for each tested supercooled van der Waals liquid, should be accompanied by the decreasing role of the density fluctuations in molecular dynamics of BMPC, PDE, and PC while the systems are approaching the glassy state. This behavior is reflected in the increasing values of the dynamic modulus $M_{p\text{-}T}$ defined by Eq. (4), which can be easily considered as the inverse of the ratio $V_T^{act}/E_p^{act}$ quantifying the decreasing role of the density fluctuations in molecular dynamics by a drop in the values $V_T^{act}/E_p^{act}$ with decreasing temperature at ambient pressure.

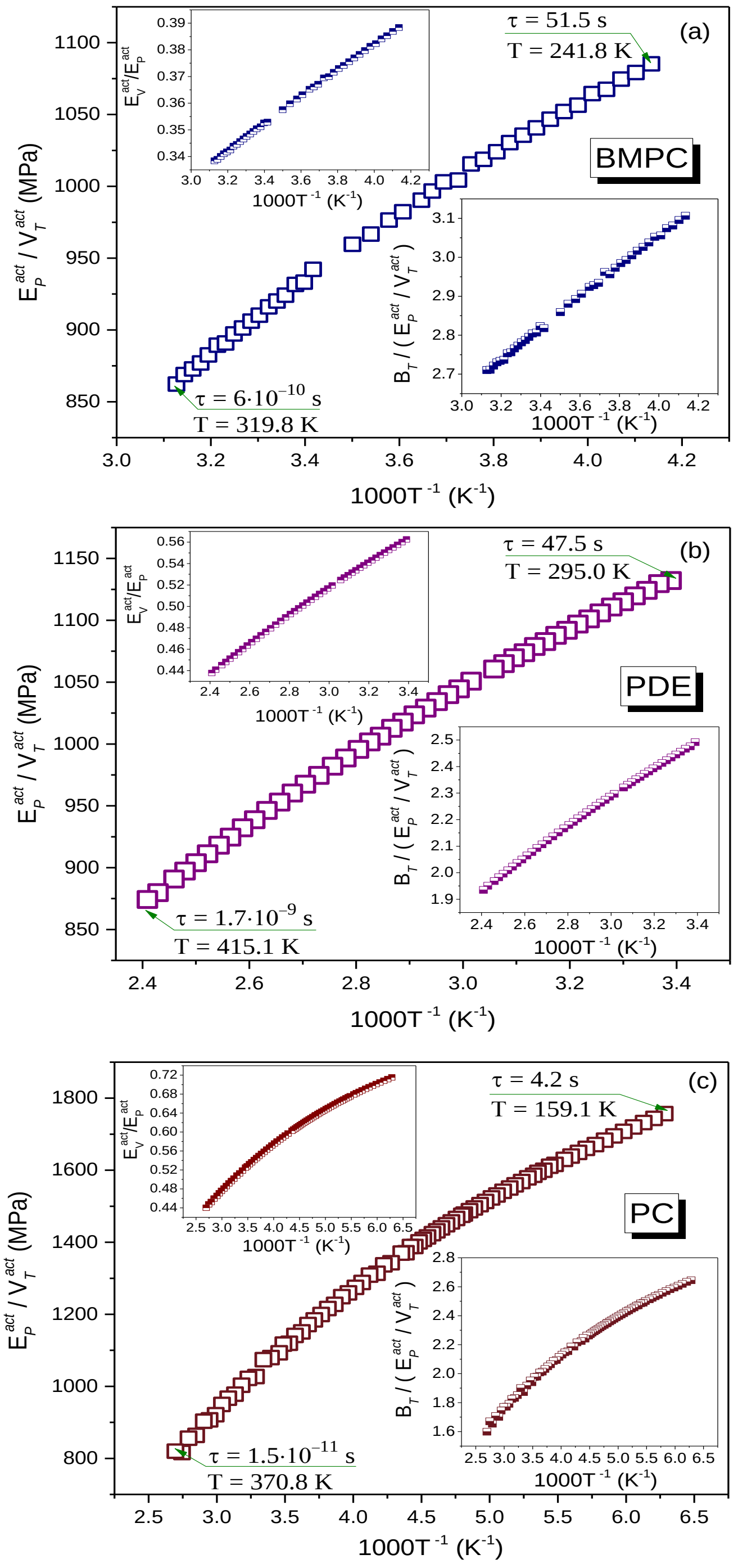

Fig. 1. Plots of the dependence of dynamic modulus $M_{p\text{-}T}$ (defined by Eq. (4) in the p-T representation) on inverse temperature for BMPC (a), PDE (b), and PC (c) at ambient pressure. The lower insets show the dependence of the ratio of thermodynamic and dynamic moduli (where the former is defined by Eq. (5)) on the inverse temperature at ambient pressure. The upper insets illustrate the dependence of the ratio of the isochoric and isobaric activation energies on the inverse temperature at ambient pressure.

The presented qualitative discussion about the interrelation between the ratio $E_V^{act}/E_p^{act}$ and the dynamic modulus, $M_{p-T} = E_p^{act}/V_T^{act}$, inclines us to find its quantitative description. Since molecular dynamics of materials belonging to the examined material group (i.e., van der Waals liquids) as well as the other large material groups (such as polymers and ionic liquids) complies with the power law density scaling law and meets its criterion given here by Eq. (6), we first investigate the sought after relation between $E_V^{act}/E_p^{act}$ and $E_p^{act}/V_T^{act}$ in the density scaling regime. Following the criterion Eq. (6), we analyze the ratios in isochronal conditions (i.e., at *τ=const*). Then, we can start from a known equation [19,20],

$$\left(\frac{E_p^{act}}{V_T^{act}}\right)_\tau = T_\tau \left(\frac{\partial p}{\partial T}\right)_\tau , \tag{7}$$

which results from the definitions of $E_p^{act}$ and $V_T^{act}$ (i.e., Eqs. (2) and (3)) considered at *τ=const*. A few years ago, we derived an equation for the glass transition temperature coefficient $dT_g/dp$ in the density scaling regime [34], which has been later generalized on the basis of the density scaling criterion (Eq. (6)) to its form valid at any *τ=const* [35],

$$\left(\frac{\partial T}{\partial p}\right)_\tau = \frac{\gamma T_\tau / B_{T_\tau}(p)}{1+\gamma T_\tau \alpha_p(T_\tau)} . \tag{8}$$

where the isobaric thermal volume expansivity can be defined as $\alpha_p = (\partial \ln V/\partial T)_p$ .

By introducing Eq. (8) into Eq. (7), we arrive at the following relationship,

$$\left(\frac{E_p^{act}}{V_T^{act}}\right)_\tau = \left(1+\gamma T_\tau \alpha_P(T_\tau)\right)\frac{B_{T_\tau}(p)}{\gamma} . \tag{9}$$

Then, taking into account a well-known representation of the ratio $E_V^{act}/E_p^{act}$ in the density scaling regime [7,10],

$$\left( \frac{E_V^{act}}{E_p^{act}} \right)_\tau = \frac{1}{1 + \gamma T_\tau \alpha_P(T_\tau)} \quad , \tag{10}$$

we find the relation between $E_V^{act} / E_p^{act}$ and $E_p^{act} / V_T^{act}$,

$$\left( \frac{E_p^{act}}{V_T^{act}} \right)_\tau = \left( \frac{E_p^{act}}{E_V^{act}} \right)_\tau \frac{B_{T_\tau}(p)}{\gamma} \quad . \tag{11}$$

In terms of Eq. (11), we can explain the obtained similar trends of the inverse temperature dependences of the ratios $B_T / M_{p-T}$ and $E_V^{act} / E_p^{act}$, where the latter can be transformed into the former by multiplying by the scaling exponent γ, i.e., $B_T / M_{p-T} = \gamma \left( E_V^{act} / E_p^{act} \right)$, at each $T$ along the isobar τ($T$) determined from the BDS measurements at ambient pressure. What is more, an interesting observation can be made from Eq. (11), which can be simplified to the following form,

$$\left( \frac{E_V^{act}}{V_T^{act}} \right)_\tau = \frac{B_{T_\tau}(p)}{\gamma} \quad , \tag{12}$$

which involves a new quantity

$$M_{V-T} = E_V^{act} / V_T^{act} \quad . \tag{13}$$

It is worth noting that Eq. (12) also has its general form, which can be derived from the definitions of the isochoric activation energy and the isothermal activation volume (Eqs. (1) and (3)) and the isothermal bulk modulus (Eq. (5)) with no density scaling limitations. The following transformations involving the partial derivative in Eq. (3), $V_T^{act} = RT \left( \partial \ln \tau / \partial \ln V \right)_T \left( \partial \ln V / \partial p \right)_T$, and then $V_{T_\tau}^{act} = -\left( RT \left( \partial \ln \tau / \partial (1/T) \right)_V / \left( \partial \ln V / \partial (1/T) \right)_\tau \right) \left( \partial \ln V / \partial p \right)_T$ , lead to the general relation, $V_{T_\tau}^{act} = -E_{V_\tau}^{act} B_{T_\tau}^{-1} T_\tau^{-1} \alpha_\tau^{-1}$, which can be reduced to Eq. (13) if the density scaling criterion is met with a constant scaling exponent γ for a given material, because then $\gamma = -T_\tau^{-1} \alpha_\tau^{-1}$ from Eq. (5), where the isochronal thermal volume expansivity has been defined as $\alpha_\tau = \left( \partial \ln V / \partial T \right)_\tau$.

It should be emphasized that the new quantity $M_{V\text{-}T}$ defined in Eq. (13), which is the ratio of the isochoric activation energy $E_V^{act}$ and the activation volume $V_T^{act}$, can be also considered as a dynamic modulus, but determined by using *isolines* in another representation (i.e., in the V-T one due to employing an isochore and an isotherm at their intersection point in the definition of $M_{V\text{-}T}$) than that suggested by Ingram et al. in the p-T representation (i.e., $M_{p\text{-}T}$ in Eq. (4)). In this way, the dynamic modulus $M_{V\text{-}T}$ straightforwardly quantifies the relative contribution of temperature and density fluctuations to molecular dynamics, while the dynamic modulus $M_{p\text{-}T}$ is an indirect measure of the relative contribution of temperature fluctuations to the overall fluctuations of $T$ and $\rho$ in molecular dynamics, because the inverse of $M_{p\text{-}T}$ quantifies the relative contribution of density fluctuations with respect to the overall temperature-density fluctuations. A question arises as to *whether or not the definition of the dynamic modulus transformed from the p-T representation to the V-T one gives us an advantage in exploring relevant invariants and characteristics of the molecular dynamics and thermodynamics of supercooled liquids*.

To answer the question, we verify the temperature dependences of the dynamic modulus $M_{V\text{-}T}$ for the selected van der Waals liquids at ambient pressure by exploiting the values of $E_V^{act}$ and $V_T^{act}$ earlier generated with the assistance of Eq. (6). As can be seen from Fig. 2, the values of $M_{V\text{-}T}$ increase with decreasing temperature even more than those determined for $M_{p\text{-}T}$ at ambient pressure for each tested supercooled liquid, i.e., the values of $M_{V\text{-}T}$ change respectively by about 44%, 66%, and 248% for BMPC, PDE, and PC in the same ranges of $T$ and $\tau$ at $p$=0.1MPa as those earlier considered for $M_{p\text{-}T}$. Thus, the dynamic modulus $M_{V\text{-}T}$ is also not any invariant parameter for a given material.

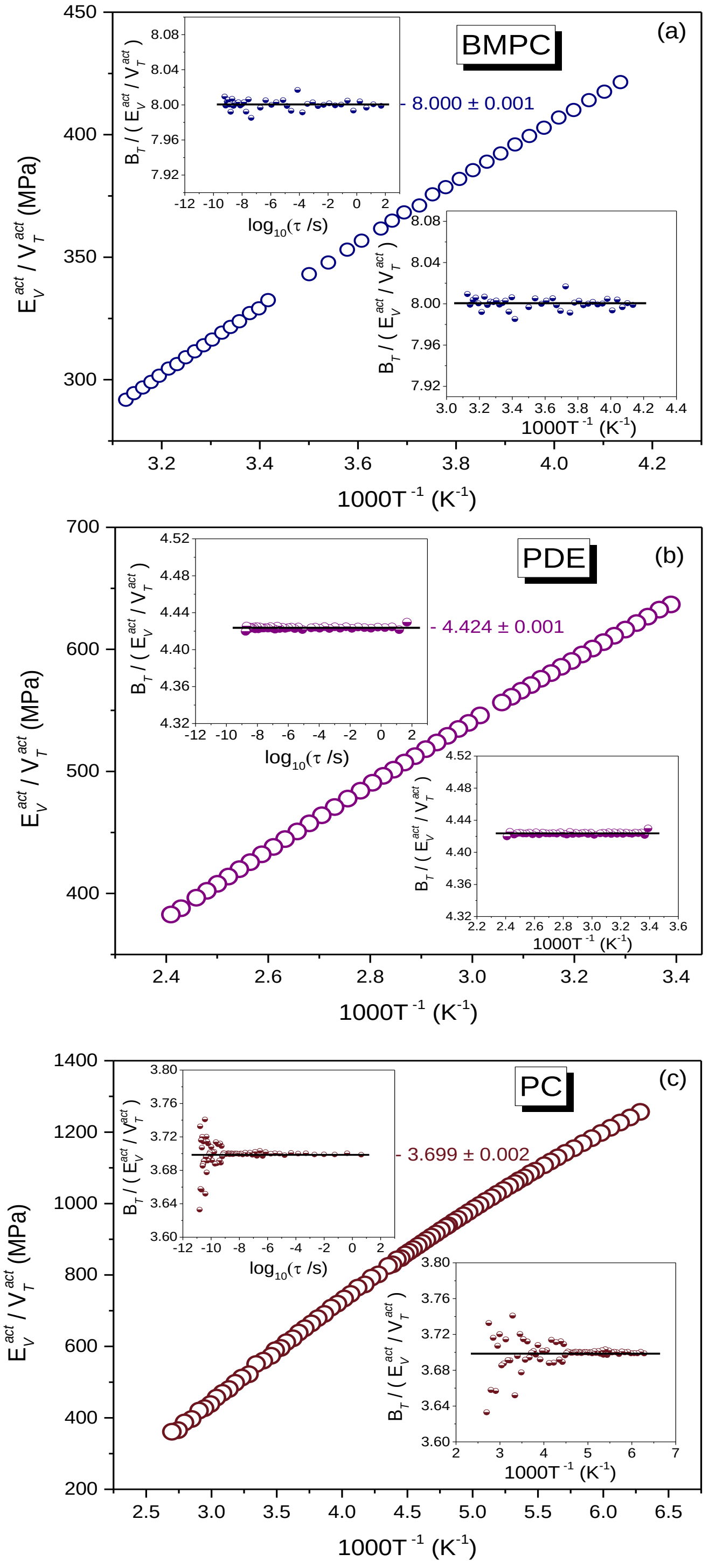

Fig. 2. Plots of the dependence of dynamic modulus $M_{V\text{-}T}$ (defined by Eq. (13) in the V-T representation) on inverse temperature for BMPC (a), PDE (b), and PC (c) at ambient pressure. The insets illustrate the dependences of the ratio of thermodynamic and dynamic moduli (defined respectively by Eqs. (5) and (13)) on the inverse temperature at ambient pressure (shown in the lower ones) and the structural relaxation times (shown in the upper ones).

In the next step, we investigate the temperature dependences of the ratio of the thermodynamic and dynamic modulus replacing the dynamic modulus $M_{p\text{-}T}$ with $M_{V\text{-}T}$. Results of the analysis are presented in the lower insets in Fig. 2, which are compared with the dependences of $B_T/M_{V\text{-}T}$ on the structural relaxation times τ in the upper insets in Fig. 2 for each tested supercooled van der Waals liquid. The outcome of the analyses is very promising, because we have found that the ratio of the thermodynamic and dynamic moduli $B_T/M_{V\text{-}T}$ can be an invariant parameter for each examined material to a very good approximation in the very broad ranges of $T$ and $\tau$ in contrast to the ratio $B_T/M_{p\text{-}T}$ earlier suggested as a material constant, which continuously varies in the wide temperature ranges (as can be easily seen by comparing the lower insets in Figs. 1 and 2 ). It should be stressed that the achieved quality of the invariance of the ratio $B_T/M_{V\text{-}T}$ is very high in the whole considered ranges of $T$ and $\tau$, which results in very small relative standard determination errors for the mean values of $B_T/M_{V\text{-}T}$, which equal only 0.013%, 0.023%, and 0.054% for BMPC, PDE, and PC, respectively. This important conclusion remains valid independent of a slightly wider distribution of the values $B_T/M_{V\text{-}T}$, which is observed for PC at the high temperatures that induce very fast molecular dynamics characterized by extremely short structural relaxation times for this material, i.e., $\tau$ < 1ns, which have been not explored experimentally in case of BMPC and PDE (see the upper insets in Fig. 2). Moreover, our thorough analysis confirms that the invariant value of the dynamic and thermodynamic moduli can be very accurately identified with the constant value of the scaling exponent for a given material, the molecular dynamics of which obeys the power density scaling law, i.e., $B_T/M_{V\text{-}T} = \gamma$ in the wide ranges of $T$ and $\tau$ at ambient pressure.

It is worth noting that such a prediction can be just made from Eq. (12), which has been firmly validated by our numerical analyses based on experimental data. Another implication of Eq. (12) can be a convenient method for validating the invariance of $B_T/M_{V\text{-}T}$, which is combined with evaluating the value of the scaling exponent γ. Based on Eq. (12), one can expect that the simple proportionality, $B_T \sim M_{V\text{-}T}$ , is satisfied if the power density scaling law is valid with a constant value of the scaling exponent $\gamma$ for a given material, because then, $B_T = \gamma M_{V\text{-}T}$. As can be seen in Fig. 3(a), such simple linear dependences of $B_T$ on $M_{V\text{-}T}$ have been achieved for each tested supercooled van der Waals liquid in the considered wide ranges of $T$ and $\tau$ at ambient pressure.

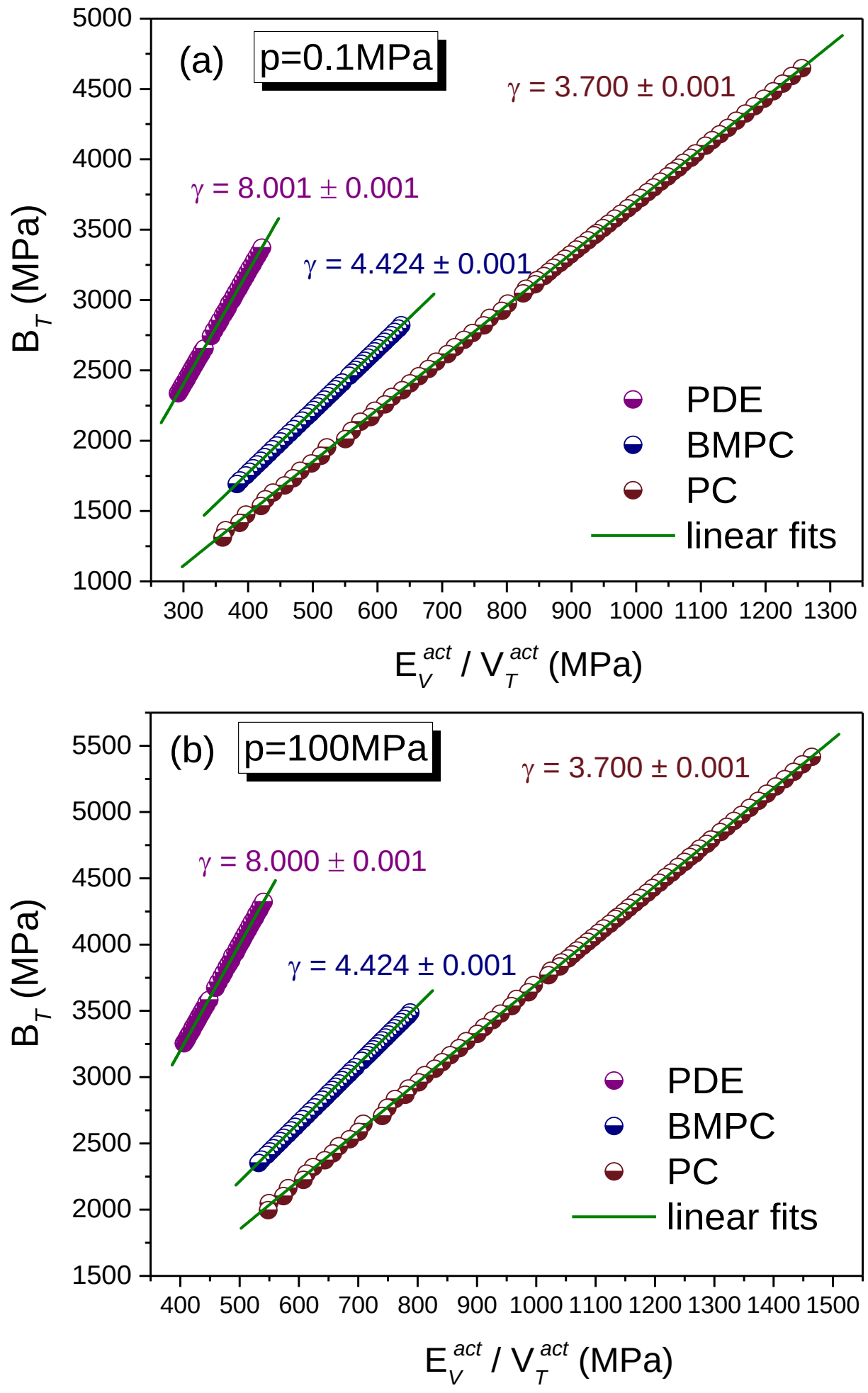

Fig. 3. Plots of linear correlations between the thermodynamic modulus $B_T$ (defined as the isothermal bulk modulus by Eq. (5)) and the dynamic modulus $M_{V\text{-}T}$ (defined in the T-V representation by Eq. (13)) for the examined supercooled van der Waals liquids in the same wide temperature ranges at (a) ambient pressure and (b) p=100MPa. The solid lines denote linear fits with the fitted slopes, which are in accord with the scaling exponents $\gamma$ for the materials.

To finally confirm the invariance of the ratio of thermodynamic and dynamic moduli $B_T/M_{V\text{-}T}$ and its equivalence to the invariant value of the density scaling exponent $\gamma$ for each tested material, we have generated additional data by using the method based on Eq. (6) to determine the values of $E_V^{act}$, $V_T^{act}$, and $B_T$ along a high pressure isobar at $p$=100MPa in the same wide ranges of $T$ and $\tau$ as those explored at ambient pressure. As a result, for each investigated material, we have obtained very high qualities of the simple linear correlation between $B_T$ and $M_{V\text{-}T}$ and the same value of the scaling exponent $\gamma$ in ambient and high pressure conditions (as shown respectively in panels (a) and (b) in Fig. 3).

Thereunder, we can claim that the ratio of dynamic and thermodynamic moduli, $M_{V\text{-}T}/B_T$, should be invariant in the entire thermodynamic space in the supercooled liquid state in case of van der Waals liquids and other materials, the molecular dynamics of which obeys the power density scaling law with a constant value of the scaling exponent for a given material. This successfully validated prediction, which can be made from Eq. (13), has significant implications for our better understanding of the linkage between molecular dynamics and thermodynamics in the wide dynamic and thermodynamic ranges (almost from the melting point to the glass transition) as well as the density scaling of molecular dynamics in viscous liquids.

## Summary and conclusions

Exploiting ambient and high pressure experimental data for selected supercooled van der Waals liquids as well as the criterion for the power density scaling law, we have verified the ideas on some material constants suggested by Ingram et al. [18-20] within the studies of ionic systems. It has turned out that neither the dynamic modulus $M_{p\text{-}T}$ (defined as the ratio of the isobaric activation energy $E_p^{act}$ and the activation volume $V_T^{act}$) nor the ratio of the thermodynamic and dynamic moduli $B_T/M_{p\text{-}T}$ (where $B_T$ is the isothermal bulk modulus) remain constant in varying thermodynamic conditions for no examined representative of van der Waals liquids (i.e., BMPC, PDE, and PC) in the wide ranges of temperatures $T$ and structural relaxation times $\tau$ at ambient pressure. The continuous increase in both the values of $M_{p\text{-}T}$ and $B_T/M_{p\text{-}T}$ with decreasing temperature has been explained in relation to the well-known measure of the contributions of temperature and density fluctuations to molecular dynamics, which is the ratio of the isochoric and isobaric activation energies $E_V^{act}/E_p^{act}$. It has been argued that the ratio $V_T^{act}/E_p^{act}$, which is the inverse of $M_{p\text{-}T}$, is a supplementary measure that quantifies the relative contribution of density fluctuations to the overall fluctuations of $T$ and $\rho$, while $E_V^{act}/E_p^{act}$ quantifies the relative contribution of temperature fluctuations to the overall temperature-density fluctuations. Therefore, the obtained increase in the values of the measures $E_V^{act}/E_p^{act}$ and $M_{p-T}\,(=E_p^{act}/V_T^{act})$ with decreasing temperature at ambient pressure consistently represent the increasing role of temperature fluctuations accompanied with the decreasing role of density fluctuations in molecular dynamics of supercooled liquids under their cooling towards the glass transition in isobaric conditions, while a significant growth is observed in the molecular packing. However, in the density scaling regime, the isobaric temperature dependence of the ratio of the thermodynamic and dynamic moduli, $B_T/M_{p\text{-}T}$, has well reproduced by the ratio $E_V^{act}/E_p^{act}$ multiplied by the density scaling exponent $\gamma$.

The aforementioned results have directed us towards the idea to transform the analysis based on the dynamic modulus from the p-T representation to the V-T one. We have defined the dynamic modulus $M_{V\text{-}T}$ as the ratio of the isochoric activation energy $E_V^{act}$ (instead of the isobaric one) and the activation volume $V_T^{act}$. In this way, we have constituted a straightforward measure of the relative contribution of temperature and density fluctuations to molecular dynamics, which is the dynamic modulus in the V-T representation in contrast to the dynamic modulus defined in the p-T representation, $M_{p\text{-}T}$, the inverse of which quantifies the relative contribution of density fluctuations with respect to the overall temperature-density fluctuations. Similarly as its counterpart in the p-T representation, the dynamic modulus $M_{V\text{-}T}$ increases with decreasing temperature in isobaric conditions, showing the increasing role of temperature fluctuations in the molecular dynamics of liquids subjected to supercooling. Nevertheless, we have established an essential application of the dynamic modulus $M_{V\text{-}T}$ defined in the V-T representation, which has enabled us to determine an invariant quantity linking thermodynamics and molecular dynamics of glass-forming liquids in the density scaling regime. By replacing the dynamic modulus $M_{p\text{-}T}$ with $M_{V\text{-}T}$, the ratio of thermodynamic and dynamic moduli has turned out to be invariant for a given material. What is more, the ratio $B_T/M_{V\text{-}T}$ has been found to be equivalent to the constant density scaling exponent $\gamma$ valid for a given material.

The latter finding is meaningful, because the ratio $B_T/M_{V\text{-}T}$ becomes a relevant parameter characterizing an interplay between thermodynamics and molecular dynamics in an invariant way for each material that complies with the power density scaling law. In this context, further investigations are certainly required to elucidate the extremely interesting result that the relative level of thermodynamic and dynamic factors quantified by the ratio $B_T/M_{V\text{-}T}$ remains unchanged in the wide thermodynamic and dynamic ranges explored in the supercooled liquid state almost from the melting point to the glass transition. Nevertheless, a preliminary attempt at taking up the challenge is simply to address to the first theoretical grounds for the power law density scaling, which suggest that the density scaling exponent is proportional to the exponent of the dominant repulsive part of the effective short-range intermolecular potential [10,43,44,45,46]. On the other hand, in consequence of our investigations, the density scaling exponent gains a robust interpretation as the material constant which indeed relates molecular dynamics and thermodynamics of viscous liquids, providing an additional argument for the concept of 'the thermodynamic scaling of molecular dynamics' as the density scaling is sometimes called.